\documentclass[aps,pra,twocolumn,nofootinbib,twocolumn,floatfix,showpacs,superscriptaddress]{revtex4-2}
\usepackage{amsmath,graphicx,epsfig,amsthm,color,bm}
\usepackage{pifont,bbding,amssymb,soul,mathrsfs,braket}
\usepackage{dcolumn}
\usepackage{siunitx}
\usepackage[hyperindex,breaklinks,colorlinks=true,citecolor=blue,urlcolor=blue]{hyperref}
\usepackage{subfigure}

\makeatletter

\newcommand{\Rmnum}[1]{\expandafter\@slowromancap\romannumeral #1@}
\makeatother

\begin{document}

\title{Broadband Quantum Photon Source in Step-Chirped Periodically Poled Lithium Niobate Waveguide}
\author{Xiao-Xu Fang}
\affiliation{School of Physics, State Key Laboratory of Crystal Materials, Shandong University, Jinan 250100, China}
\author{Guoliang Shentu}
\affiliation{Wuhan Institute of Quantum Technology, Wuhan 430000, China}
\affiliation{Shandong Guoyao Quantum Lidar Technology Company Ltd., Jinan 250101, China}
\author{He Lu}
\email{luhe@sdu.edu.cn}
\affiliation{School of Physics, State Key Laboratory of Crystal Materials, Shandong University, Jinan 250100, China}


\begin{abstract}
Broadband nonlinear optical devices play a critical role in both classical and quantum optics. Here, we design and fabricate a 6.82-mm-long step-chirped periodically poled lithium niobate~(CPPLN) waveguide on lithium niobate on insulator, which enables quasi-phase matching over a broad bandwidth for second-harmonic generation~(SHG) and spontaneous parametric down-conversion~(SPDC). The SHG achieves an average efficiency of 54.4\%/W/cm$^2$ over the first-harmonic wavelength range of 1510~nm-1620~nm, paving the way for realizing SPDC across a wide range of pump wavelengths. For SPDC, by tuning the pump wavelength to 775~nm, 780~nm, and 785~nm, we achieve broadband photon-pair generation with a maximum full bandwidth and brightness up to 99~THz~(846~nm) and 20~GHz/mW/nm, respectively. Our findings provide an efficient and experiment-friendly approach for generating broadband photon pairs, which holds significant promise for advancing applications in quantum metrology.
\end{abstract}

\maketitle

\section{\label{sec:level1}Introduction}
Spontaneous parametric down-conversion~(SPDC)~\cite{Burnham1970PRL} is a fundamental nonlinear optical process, wherein a single pump photon is converted into a pair of correlated photons. SPDC is critical to generate quantum photon sources, including heralded single photon~\cite{Hong1986PRL}, entangled photon pair~\cite{Kwiat1995PRL}, and squeezed light~\cite{Wu1986PRL}. Broadband photon pairs generated via SPDC have attracted increasing attention and exhibit significant potential for emerging quantum photonic technologies, such as quantum imaging~\cite{Lemos2014Nature, Kalashnikov2016NP, Paterova2018NJP}, frequency multiplexing~\cite{Puigibert2017PRL, Joshi2018NC}, high-dimensional encoding~\cite{Olislager2010PRA, Sheridan2010PRA} and quantum optical coherence tomography~\cite{Abouraddy2002PRA, Hayama2022OL}. Lithium niobate on insulator~(LNOI) possesses a large second-order nonlinear susceptibility~($d_{33}$=-27 pm/V) and a broad transparency window~(0.35-5.2~$\mu$m)~\cite{Weis1985APL}. Its ferroelectric property enables periodically domain engineering, i.e., periodically poling lithium niobate~(PPLN), to achieve quasi-phase matching~(QPM), which is a crucial prerequisite for efficient SPDC. Moreover, the tailored design of the waveguide structure enables the engineering of group velocity dispersion~(GVD), thereby facilitating broadband nonlinear optical processes, including SPDC~\cite{Xue2021PRA, Javid2021PRL,Fang2024OE}, sum-frequency generation~\cite{Li2017OL}, second harmonic generation~(SHG) and supercontinuum generation~\cite{Jankowski2020Optica}. However, the bandwidth of SPDC photons is quite sensitivity the wavelength of pump light~\cite{Javid2021PRL}.


Nonlinear crystals with chirped poling period provide an alternative solution, where varying poling periods are designed to satisfy QPM across a range of wavelengths---thereby enabling the generation of broadband SPDC. The step-chirped periodical poling has been demonstrated in various nonlinear crystals for SPDC, including LN~\cite{Yadav2022OL, Roeder2024arXiv}, potassium titanyl phosphate~(KTP)~\cite{Shaked2014NJP} and stoichiometric lithium tantalate~(SLT)~\cite{Nasr2008PRL,Tanaka2012OE,Cao2021OE,Hojo2021SR,Cao2023OE,Tashima2024Optica}. Specifically, the broadband property of SPDC in chirped PPLN~(CPPLN) has been theoretically investigated~\cite{Harris2007PRL,Zhu2025APL}, and experimentally demonstrated using zinc~(Zn) in-diffused PPLN~\cite{Yadav2022OL} and titanium~(Ti) in-diffused PPLN~\cite{Roeder2024arXiv}, respectively. The waveguides fabricated via in-diffused technique are with larger optical modes, resulting weak optical confinement. In contrast, nanoscale waveguide structures on LNOI enable strong confinement the interacting waves, thereby significantly enhancing the SPDC efficiency. On the other side, the chirped PPLN enables broadband SHG~\cite{Chen2014Light,Chen2021Research}, alleviating the requirement of pump wavelength in SPDC.      

In this Letter, we design and fabricate a 6.82-mm-long step-chirped PPLN waveguide on LNOI, which simultaneously achieves broadband SHG and SPDC. The SHG phase-matching bandwidth exceeds 110~nm, with an average efficiency of 54.4\%/W/cm$^2$ across 1510~nm to 1620~nm. By setting the pump wavelength at 775~nm, 780~nm, and 785~nm, we observe photon-pair generation with a maximum full bandwidth and brightness up to 99~THz~(846~nm) and 20~GHz/mW/nm, respectively.

\section{Design and fabrication of the step-chirped PPLN waveguide}
\begin{figure*}
\centering
	\includegraphics[width=\linewidth]{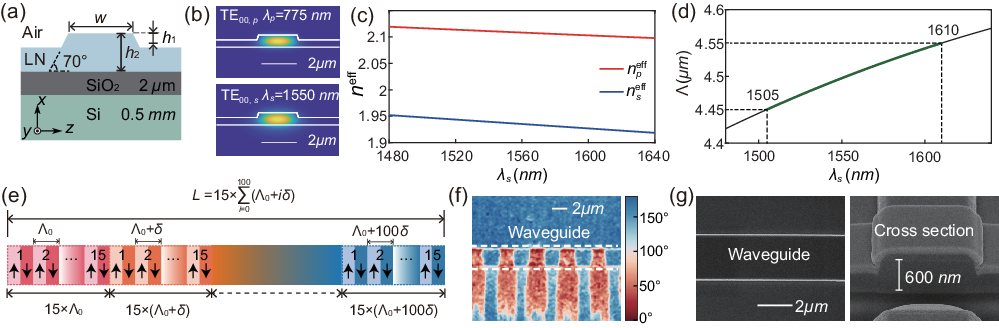}
\caption{(a) Cross-section of the ridge waveguide fabricated on a $h_2=600$~nm thick $x$-cut LNOI wafer. (b) The mode profiles of TE$_{00}$ of $\lambda_p=775$~nm~(top) and $\lambda_s=1550$~nm~(bottom). (c) Simulated effective refractive indices $n^\text{eff}$ of TE$_{00,s}$~(blue) and TE$_{00,p}$~(red) with $\lambda_s\in[1480~\text{nm}, 1640~\text{nm}]$. (d) Poling periods to satisfy QPM between $\lambda_{p}$ and $\lambda_{s}$ with $\lambda_s\in[1480~\text{nm}, 1640~\text{nm}]$. (d) Design of the step-chirped PPLN. Upward~(downward) arrows indicate the sign of the nonlinear susceptibility $+(-)$. (e) PFM phase image of the periodical poling region. (f) SEM images of the fabricated waveguide.}
\label{Fig:1}
\end{figure*}

As shown in Figure~\ref{Fig:1}~(a), the LNOI sample in our experiment consists of 600-nm-thick x-cut LN thin film bonded to 0.5-mm-think Si substrate with 2-$\mu m$-think SiO$_2$~(NANOLN Inc.). The cross-section of waveguide we designed is with top width of $w=2200$~nm, etching depth of $h_1=240$~nm and a sidewall angle of $70^{\circ}$. In SPDC, a short-wavelength pump photon~($p$) is spontaneously split into a pair of longer-wavelength photons, i.e., signal photon~($s$) and idler photon~($i$). In this three-wave mixing precess, energy conservation must be satisfied $\omega_p=\omega_s+\omega_i$ with $\omega$ being the frequency of the respective photons. More importantly, phase-matching condition $\Delta k=0$ for the three interacting waves must be satisfied to enable efficient SPDC. In QPM, the phase mismatch between the pump light and signal/idler photon can be compensated by reversing the sign of the nonlinear susceptibility at regular intervals of $\Lambda/2$, i.e.,
\begin{equation}
\Delta k=k_p-k_s-k_i-\frac{2m\pi}{\Lambda}=0,
\end{equation} 
where $k_x=2\pi n^\text{eff}_x/\lambda_{x}~(x=p,s,i)$ is wave vector and $n^\text{eff}_x$ is the effective refractive index of the pump, signal, and idler photons, respectively. Consider the degenerate case $\lambda_s=\lambda_i=\lambda_p/2$, the poling period is 
\begin{equation}\label{Eq:period}
\Lambda = \frac{\lambda_p}{n^\text{eff}_{p}-n^\text{eff}_{s}}
\end{equation}
To utilize the largest second-order nonlinear susceptibility of LN, we consider the SPDC based on type-0 phase-matching, i.e., $\text{TE}_{00,p}\to \text{TE}_{00,s}+\text{TE}_{00,i}$ with $\text{TE}_{00,x}~(x=p,s,i)$ being the fundamental transverse-electric~(TE) mode for pump, signal and idler respectively~(as shown in Figure~\ref{Fig:1}~(b). Given the cross section in Figure~\ref{Fig:1}~(a), we simulated the effective refractive index $n^\text{eff}$ of $\text{TE}_{00,p}$ and $\text{TE}_{00,s}$ for signal wavelength $\lambda_s$ ranging from 1480~nm to 1640~nm, and the results are shown in Figure~\ref{Fig:1}~(c). The corresponding poling periods are then calculated according the Equation~\ref{Eq:period}, as shown in Figure ~\ref{Fig:1}~(d). The step-chirped poling period under consideration ranges from 4.45~$\mu$m to 4.55~$\mu$m~(green shaded region in Figure ~\ref{Fig:1}~(d)), corresponding to $\lambda_s$ from 1505~nm to 1610~nm. As shown in Figure~\ref{Fig:1}~(e), step-chirped periodical poling is divided into 101 sections, each containing 15 fixed periods $\Lambda$ increase from 4.45~$\mu$m to 4.55~$\mu$m with step size of $\delta=1$~nm. We denote $\Lambda_0=4.45~\mu$m, and the the total length of the CPPLN waveguide is $L=15\times\sum_{i=0}^{100}(\Lambda_0+i\delta)=6.8175~\text{mm}$.

To fabricate the designed CPPLN waveguide, we first create chromium~(Cr) electrodes on the surface of LNOI via electron beam lithography (EBL), followed by deposition and lift-off processes. Next, the pulsed voltage is applied to the Cr electrodes to induce domain poling. Finally, the waveguide pattern is defined using EBL, and the waveguide is fabricated via reactive ion etching. The fabrication process is similar to Ref.~\cite{Fang2024OE}. As shown in Figure~\ref{Fig:1}~(f), the periodical poling region is characterized by piezoresponse force microscopy~(PFM), where the red areas correspond to the inverted domains and the edge of waveguide is marked with white dashed lines. Scanning electron microscope~(SEM) images of the fabricated waveguide are shown in Figure~\ref{Fig:1}~(g). 

\section{Characterization of the step-chirped PPLN waveguide}
\subsection{Results of SHG}
\begin{figure}
\centering
	\includegraphics[width=\linewidth]{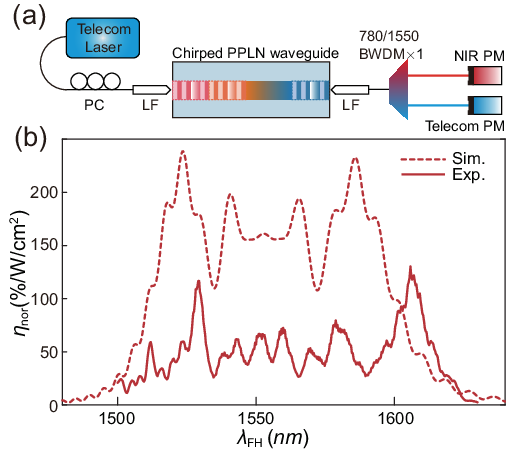}
\caption{(a) The experimental setup for characterizing the SHG process. (b) Experimental~(red solid line) and simulation~(red dashed line) results of the normalized SHG efficiency.}
\label{Fig:2}
\end{figure}

Before testing the SPDC, we characterize the classical SHG efficiency of the fabricated CPPLN waveguide with the setup in Figure~\ref{Fig:2}~(a). The first-harmonic~(FH) light in the telecom band is provided by a continuous-wave~(CW) tunable laser~(Santec, TSL550). A polarization controller~(PC) is used to ensure the FH light maintained TE polarization, which is then coupled into CPPLN waveguide via a lensed fiber~(LF). The generated second-harmonic~(SH) light, along with the residual FH light, is coupled out through another LF and subsequently separated by a 775/1550~nm band wavelength division multiplexer~(BWDM). The power intensities of FH and SH lights~(denoted as $P_\text{FH}$ and $P_\text{SH}$, respectively), are recorded with a telecom power meter~(PM) and a near infrared~(NIR) PM, respectively. The normalized SHG efficiency $\eta_\text{nor}$ is calculated by
\begin{equation}
\eta_{\text{nor}}=\frac{P_\text{SH}/\eta_\text{SH}}{(P_\text{FH}/\eta_\text{FH})^2 \cdot L^2},
\end{equation}
where $\eta_\text{FH}=19.28\%$ and $\eta_\text{SH}=7.91\%$ are the coupling efficiency from fiber to waveguide, for FH and SH lights respectively. By sweeping the wavelength of the FH light, we obtain the SHG spectrum~(in terms of $\eta_{\text{nor}}$) as shown by the solid red line in Figure~\ref{Fig:2}~(b). Over the FH wavelength range of 1510~nm to 1620~nm, we observe an average $\eta_{\text{nor}}$ of 54.4\%/W/cm$^2$.  

We also calculate the theoretical value of $\eta_{\text{nor}}^\text{th}$, given by \cite{Wu2022OL, Zhang2022OME}
\begin{equation}
\eta_{\text{nor}}^\text{th} = \frac{8 \pi^2}{\epsilon_0 c n_1^2 n_2 \lambda_{\omega}^2 } \frac{ d_{\mathrm{eff}}^2}{S_{\mathrm{eff}}} G^2(\Delta k),
\end{equation}
where
\begin{align}
&S_{\mathrm{eff}} =\frac{\left[ \iint{E_{z,\text{FH}}^{2}}(x,z)dxdz \right] ^2\iint{E_{z,\text{SH}}^{2}}(x,z)dxdz}{\left[ \iint{d}(x,z)E_{z,\text{FH}}^{2}(x,z)E_{z,\text{SH}}(x,z)dxdz \right] ^2}, \\
&G^2(\Delta k) = \left|\frac{1}{L}\int_0^L{d}(z) e^{ -i\Delta k z } dz \right|^2.
\end{align}
Here, $\epsilon_0$ and $c$ are the permittivity and the speed of light in vacuum, respectively. $n_\text{FH}$ and $n_\text{SH}$ are the effective refractive indices of the FH and SH lights, respectively. $E_{z,\text{FH}}(x,z)$ and $E_{z,\text{SH}}(x,z)$ are the electric fields of TE$_{00}$ modes along $z$ direction of FH and SH lights, respectively. The effective nonlinear susceptibility is $d_\text{eff}=d_{33}=-27$~pm/V. The results of $\eta_{\text{nor}}^\text{th}$ are shown with red dashed line in Figure~\ref{Fig:2}~(b), which yields an average efficiency of 165\%/W/cm$^2$ from 1510~nm to 1600~nm. The discrepancy between the experimental and simulation results is primarily attributed to fabrication imperfections. Specifically, structural imperfections in the waveguide induce a shift in the overall spectrum, while non-ideal domain duty cycles lead to a reduction in SHG efficiency.

\begin{figure*}
\centering
	\includegraphics[width=\linewidth]{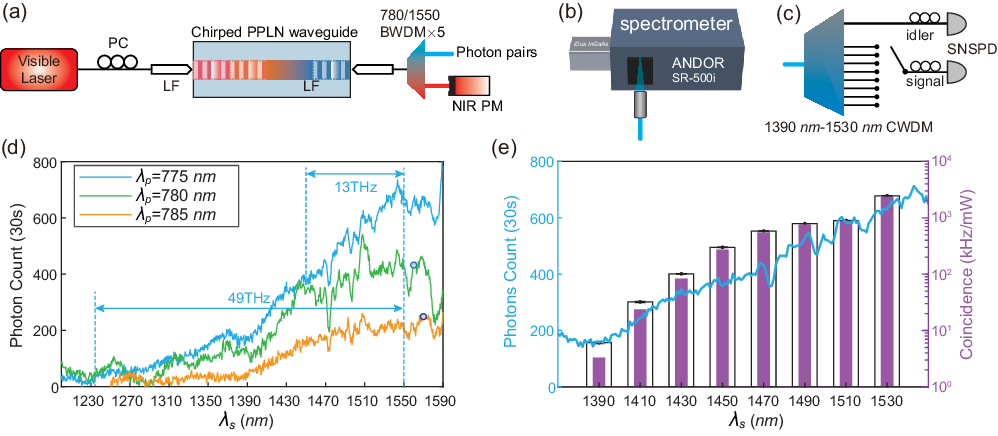}
\caption{(a) The setup for pumping CPPLN waveguide to generate photon pairs. (b) The setup for measuring the spectrum of the signal photon. (c) The setup to measure the coincidence of signal and idler photons, where the signal photon is filtered across different channels. (d) Measured spectrum of signal photons with the pump wavelength set at 775~nm, 780~nm and 785~nm, respectively. (e) Coincidences of photon pairs by setting the wavelength of pump light at 775~nm. The blue line represents the spectrum measured by spectrometer~(shown in~(d)). The black frames represent the raw coincidence counts, while purple bars represent the coincidence counts after subtracting the accidental coincidences.}
\label{Fig:3}
\end{figure*}

\subsection{SPDC}
\subsubsection{The bandwidth of SPDC}
The setup to generate photon pairs is shown in~Figure~\ref{Fig:3}~(a). A tunable CW NIR laser is used to pump the CPPLN waveguide, generating signal and idler photons. At the output, the pump light is separated from the signal/idler photons using 5 BWDMs at the output. The spectrum of the SPDC is measured by coupling the signal and idler photons into an infrared spectrometer~(ANDOR, Shamrock SR-500i) equipped with an InGaAs camera~(ANDOR iDus DU490A-1.7), as shown in Figure~\ref{Fig:3}~(b). First, the pump laser wavelength was set to 775 nm, and the corresponding SPDC spectrum is presented as the blue line in Figure~\ref{Fig:3}~(d). The measured spectrum is truncated at 1590~nm, which is attributed to the limited detection efficiency of the InGaAs camera for wavelengths longer than this value. Hereafter, the photon with wavelength shorter~(longer) than the spectral-center wavelength is referred as signal~(idler). 

To estimate the bandwidth of signal photon, we only consider the wavelengths with photon counts at least three times higher than the standard deviation of the background noise~\cite{Cao2021OE}. Based on this criterion, the spectrum of signal photon ranges from 1235~nm to 1550~nm, corresponding to a full bandwidth of 49~THz and a 3-dB bandwidth of 13~THz. Using the energy conservation relation $\omega_p=\omega_s+\omega_i$, the full bandwidth and 3-dB bandwidth of SPDC spectrum are estimated to be 99~THz~(846~nm) and 26~THz~(206~nm), respectively. We also measure the signal photon spectra with pump wavelengths set at 780~nm and 785~nm, which are shown with green and orange lines in Figure~\ref{Fig:3}~(d), respectively. The full bandwidth~(3-dB bandwidth) of the SPDC process was estimated to be 78~THz~(34 THz) and 90~THz~(33~THz) for pump wavelengths of 780 nm and 785 nm, respectively.

The bandwidth of SPDC is also characterized by analyzing the coincidence rate of signal and idler photons. To this end, the pump light wavelength was set to 775 nm, and the signal and idler photons were separated using a coarse wavelength division multiplexer~(CWDM), as illustrated in Figure~\ref{Fig:3}~(c). Specifically, the CWDM routes the idler photon to a dedicated channel (denoted as $i$), while splitting the signal photon into 8 discrete channels spanning the wavelength range of 1390~nm to 1530~nm, with a channel spacing of 20~nm and a 3-dB bandwidth of 18~nm per channel. The signal and idler photons are detected by superconducting nanowire single-photon detectors~(SPSNDs), and recorded by a time-correlated single-photon counting~(TCSPC) system. The results of coincidence rate are shown in Figure~\ref{Fig:3}~(e), where the black frames represent the raw coincidence counts and purple bars represent the coincidence counts after subtracting the accidental coincidences. We observe the maximal coincidence rate of 2.46~MHz/mW at $\lambda_s=1530$~nm and minimal coincidence rate of 6.08~kHz/mW at $\lambda_s=1390$~nm. The trend of the coincidence rate is in good agreement with the spectrum of the signal photon.

\subsubsection{Brightness and temporal correlation of photon pairs}
\begin{figure*}{ht!}
\centering
	\includegraphics{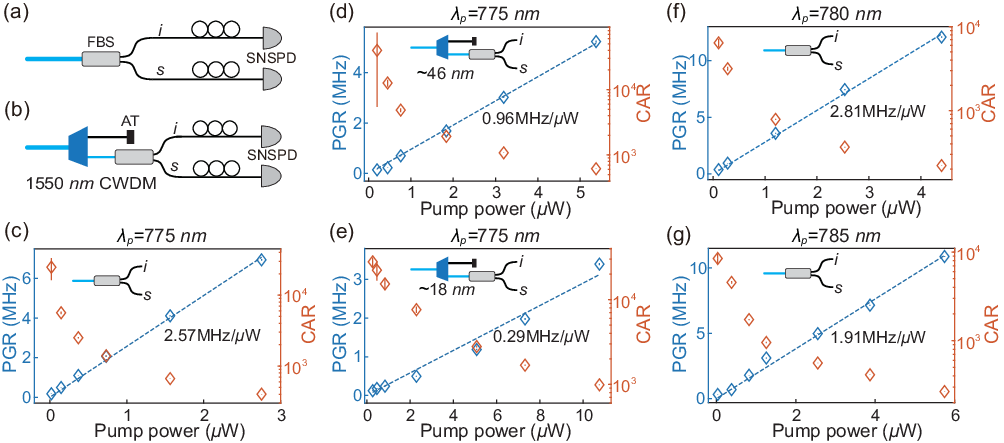}
\caption{(a) Setup for measuring the pair generation rate~(PGR) using a fiber beam splitter~(FBS). (b) Setup for measuring the PGR using a coarse wavelength division multiplexer~(CWDM) and an FBS.(c)–(g) Measured PGR and coincidence-to-accidental ratio~(CAR) under different pump wavelengths and measurement configurations.}\label{Fig:4}
\end{figure*}

To characterize the brightness of photon-pair generation, we use a 50:50 fiber beam splitter~(FBS) to split the signal and idler photons as illustrated in~Figure~\ref{Fig:4}~(a). After passing through the FBS, the signal and idler photons are detected by SNSPDs and recorded by the TCSPC system. We denote the count rates of signal and idler photons as $C_s$ and $C_i$, respectively, and the coincidence count rate as $C_{si}$. The pair generation rate~(PGR) is then calculated by $C_iC_s/(2C_{si})$, where a factor of 2 in the denominator accounts for the FBS in the experimental configuration for detection. As shown in~Figure~\ref{Fig:4}~(c), we measure the PGR of CPPLN pumped by $\lambda_p=775$ at different powers. By linear fitting of the PGR, we obtain an on-chip PGR of 2.57~MHz/$\mu$W. By shifting the wavelength of pump light to 780~nm and 785~nm, we observe PGR of 2.81~MHz/$\mu$W and 1.9~MHz/$\mu$W as shown in Figures~\ref{Fig:4}~(f) and (g) respectively, which is slightly smaller than that of the waveguide pumped at 775~nm.    

Considering that the PGR might be underestimated due to the limited detection wavelength range of the SNSPDs, we use CWDMs centered at 1550~nm to filter out the signal and idler photons~(as shown in Figure~\ref{Fig:4}~(b)). Two CWDMs with bandwidths of 46~nm and 18~nm are employed in our experiment, and the corresponding PGRs at different pump powers after filtering are shown in Figures~\ref{Fig:4}~(d) and (e). According to the fitting, we estimate the PGR to be 0.96~MHz/$\mu$W and 0.29~MHz/$\mu$W respectively. Dividing the PRG by the respective bandwidths~(46~nm and 18~nm) gives the brightness of 20~GHz/mW/nm and 16~GHz/mW/nm. 

The temporal correlation of signal and idler photons is characterized by the coincidence-to-accidental ratio~(CAR), which is obtained by measuring the $C_{si}$ by introducing the time delay between signal and idler photons. The CAR is then calculated by 
\begin{equation}
\text{CAR}=\frac{C_{si}(0)-C_{si}(\infty)}{C_{si}(\infty)},
\end{equation}
where $C_{si}(0)$ is the coincidence count rate at time delay of $t=0$ and $C_{si}(\infty)$ is the accidental coincidence count rate at a time delay far from $t=0$. As shown with red diamonds in Figures~\ref{Fig:4}~(c)-(g), the CAR value decreases with increasing PGR because of higher-order excitation in SPDC, which correspondingly reduces the temporal correlation.

\section{Conclusion}
In conclusion, we design and fabricate a step-chirped PPLN waveguide on LNOI platform, demonstrating its capability for efficient broadband SHG and SPDC---two key nonlinear optical processes for classical and quantum photonic applications. For SHG, the CPPLN waveguide achieves a QPM bandwidth exceeding 110~nm and an average normalized efficiency of 54.4\%/W/cm$^2$ across 1510~nm-1620~nm. For SPDC, broadband photon-pair generation is realized with pump wavelengths of 775~nm, 780~nm, and 785~nm, yielding full bandwidths up to 99~THz~(846~nm) and brightness up to 20~GHz/mW/nm. Photon pairs with the broad bandwidth and high brightness is able to achieve Hong-Ou-Mandel~(HOM) interference with narrow dip~\cite{Hong1987PRL}, which would benefit quantum metrology applications~\cite{Lyons2018SA,Guo2023OE,Jin2024PQE}.

\bibliography{LN_period.bib}

\begin{thebibliography}{38}%
\makeatletter
\providecommand \@ifxundefined [1]{%
 \@ifx{#1\undefined}
}%
\providecommand \@ifnum [1]{%
 \ifnum #1\expandafter \@firstoftwo
 \else \expandafter \@secondoftwo
 \fi
}%
\providecommand \@ifx [1]{%
 \ifx #1\expandafter \@firstoftwo
 \else \expandafter \@secondoftwo
 \fi
}%
\providecommand \natexlab [1]{#1}%
\providecommand \enquote  [1]{``#1''}%
\providecommand \bibnamefont  [1]{#1}%
\providecommand \bibfnamefont [1]{#1}%
\providecommand \citenamefont [1]{#1}%
\providecommand \href@noop [0]{\@secondoftwo}%
\providecommand \href [0]{\begingroup \@sanitize@url \@href}%
\providecommand \@href[1]{\@@startlink{#1}\@@href}%
\providecommand \@@href[1]{\endgroup#1\@@endlink}%
\providecommand \@sanitize@url [0]{\catcode `\\12\catcode `\$12\catcode
  `\&12\catcode `\#12\catcode `\^12\catcode `\_12\catcode `\%12\relax}%
\providecommand \@@startlink[1]{}%
\providecommand \@@endlink[0]{}%
\providecommand \url  [0]{\begingroup\@sanitize@url \@url }%
\providecommand \@url [1]{\endgroup\@href {#1}{\urlprefix }}%
\providecommand \urlprefix  [0]{URL }%
\providecommand \Eprint [0]{\href }%
\providecommand \doibase [0]{https://doi.org/}%
\providecommand \selectlanguage [0]{\@gobble}%
\providecommand \bibinfo  [0]{\@secondoftwo}%
\providecommand \bibfield  [0]{\@secondoftwo}%
\providecommand \translation [1]{[#1]}%
\providecommand \BibitemOpen [0]{}%
\providecommand \bibitemStop [0]{}%
\providecommand \bibitemNoStop [0]{.\EOS\space}%
\providecommand \EOS [0]{\spacefactor3000\relax}%
\providecommand \BibitemShut  [1]{\csname bibitem#1\endcsname}%
\let\auto@bib@innerbib\@empty
\bibitem [{\citenamefont {Burnham}\ and\ \citenamefont
  {Weinberg}(1970)}]{Burnham1970PRL}%
  \BibitemOpen
  \bibfield  {author} {\bibinfo {author} {\bibfnamefont {D.~C.}\ \bibnamefont
  {Burnham}}\ and\ \bibinfo {author} {\bibfnamefont {D.~L.}\ \bibnamefont
  {Weinberg}},\ }\bibfield  {title} {\bibinfo {title} {Observation of
  simultaneity in parametric production of optical photon pairs},\ }\href
  {https://doi.org/10.1103/PhysRevLett.25.84} {\bibfield  {journal} {\bibinfo
  {journal} {Phys. Rev. Lett.}\ }\textbf {\bibinfo {volume} {25}},\ \bibinfo
  {pages} {84} (\bibinfo {year} {1970})}\BibitemShut {NoStop}%
\bibitem [{\citenamefont {Hong}\ and\ \citenamefont
  {Mandel}(1986)}]{Hong1986PRL}%
  \BibitemOpen
  \bibfield  {author} {\bibinfo {author} {\bibfnamefont {C.~K.}\ \bibnamefont
  {Hong}}\ and\ \bibinfo {author} {\bibfnamefont {L.}~\bibnamefont {Mandel}},\
  }\bibfield  {title} {\bibinfo {title} {Experimental realization of a
  localized one-photon state},\ }\href
  {https://doi.org/10.1103/PhysRevLett.56.58} {\bibfield  {journal} {\bibinfo
  {journal} {Physical Review Letters}\ }\textbf {\bibinfo {volume} {56}},\
  \bibinfo {pages} {58} (\bibinfo {year} {1986})}\BibitemShut {NoStop}%
\bibitem [{\citenamefont {Kwiat}\ \emph {et~al.}(1995)\citenamefont {Kwiat},
  \citenamefont {Mattle}, \citenamefont {Weinfurter}, \citenamefont
  {Zeilinger}, \citenamefont {Sergienko},\ and\ \citenamefont
  {Shih}}]{Kwiat1995PRL}%
  \BibitemOpen
  \bibfield  {author} {\bibinfo {author} {\bibfnamefont {P.~G.}\ \bibnamefont
  {Kwiat}}, \bibinfo {author} {\bibfnamefont {K.}~\bibnamefont {Mattle}},
  \bibinfo {author} {\bibfnamefont {H.}~\bibnamefont {Weinfurter}}, \bibinfo
  {author} {\bibfnamefont {A.}~\bibnamefont {Zeilinger}}, \bibinfo {author}
  {\bibfnamefont {A.~V.}\ \bibnamefont {Sergienko}},\ and\ \bibinfo {author}
  {\bibfnamefont {Y.}~\bibnamefont {Shih}},\ }\bibfield  {title} {\bibinfo
  {title} {New {{High-Intensity Source}} of {{Polarization-Entangled Photon
  Pairs}}},\ }\href {https://doi.org/10.1103/PhysRevLett.75.4337} {\bibfield
  {journal} {\bibinfo  {journal} {Physical Review Letters}\ }\textbf {\bibinfo
  {volume} {75}},\ \bibinfo {pages} {4337} (\bibinfo {year}
  {1995})}\BibitemShut {NoStop}%
\bibitem [{\citenamefont {Wu}\ \emph {et~al.}(1986)\citenamefont {Wu},
  \citenamefont {Kimble}, \citenamefont {Hall},\ and\ \citenamefont
  {Wu}}]{Wu1986PRL}%
  \BibitemOpen
  \bibfield  {author} {\bibinfo {author} {\bibfnamefont {L.-A.}\ \bibnamefont
  {Wu}}, \bibinfo {author} {\bibfnamefont {H.~J.}\ \bibnamefont {Kimble}},
  \bibinfo {author} {\bibfnamefont {J.~L.}\ \bibnamefont {Hall}},\ and\
  \bibinfo {author} {\bibfnamefont {H.}~\bibnamefont {Wu}},\ }\bibfield
  {title} {\bibinfo {title} {Generation of {{Squeezed States}} by {{Parametric
  Down Conversion}}},\ }\href {https://doi.org/10.1103/PhysRevLett.57.2520}
  {\bibfield  {journal} {\bibinfo  {journal} {Physical Review Letters}\
  }\textbf {\bibinfo {volume} {57}},\ \bibinfo {pages} {2520} (\bibinfo {year}
  {1986})}\BibitemShut {NoStop}%
\bibitem [{\citenamefont {Lemos}\ \emph {et~al.}(2014)\citenamefont {Lemos},
  \citenamefont {Borish}, \citenamefont {Cole}, \citenamefont {Ramelow},
  \citenamefont {Lapkiewicz},\ and\ \citenamefont
  {Zeilinger}}]{Lemos2014Nature}%
  \BibitemOpen
  \bibfield  {author} {\bibinfo {author} {\bibfnamefont {G.~B.}\ \bibnamefont
  {Lemos}}, \bibinfo {author} {\bibfnamefont {V.}~\bibnamefont {Borish}},
  \bibinfo {author} {\bibfnamefont {G.~D.}\ \bibnamefont {Cole}}, \bibinfo
  {author} {\bibfnamefont {S.}~\bibnamefont {Ramelow}}, \bibinfo {author}
  {\bibfnamefont {R.}~\bibnamefont {Lapkiewicz}},\ and\ \bibinfo {author}
  {\bibfnamefont {A.}~\bibnamefont {Zeilinger}},\ }\bibfield  {title} {\bibinfo
  {title} {Quantum imaging with undetected photons},\ }\href
  {https://doi.org/10.1038/nature13586} {\bibfield  {journal} {\bibinfo
  {journal} {Nature}\ }\textbf {\bibinfo {volume} {512}},\ \bibinfo {pages}
  {409} (\bibinfo {year} {2014})}\BibitemShut {NoStop}%
\bibitem [{\citenamefont {Kalashnikov}\ \emph {et~al.}(2016)\citenamefont
  {Kalashnikov}, \citenamefont {Paterova}, \citenamefont {Kulik},\ and\
  \citenamefont {Krivitsky}}]{Kalashnikov2016NP}%
  \BibitemOpen
  \bibfield  {author} {\bibinfo {author} {\bibfnamefont {D.~A.}\ \bibnamefont
  {Kalashnikov}}, \bibinfo {author} {\bibfnamefont {A.~V.}\ \bibnamefont
  {Paterova}}, \bibinfo {author} {\bibfnamefont {S.~P.}\ \bibnamefont
  {Kulik}},\ and\ \bibinfo {author} {\bibfnamefont {L.~A.}\ \bibnamefont
  {Krivitsky}},\ }\bibfield  {title} {\bibinfo {title} {Infrared spectroscopy
  with visible light},\ }\href {https://doi.org/10.1038/nphoton.2015.252}
  {\bibfield  {journal} {\bibinfo  {journal} {Nature Photon}\ }\textbf
  {\bibinfo {volume} {10}},\ \bibinfo {pages} {98} (\bibinfo {year}
  {2016})}\BibitemShut {NoStop}%
\bibitem [{\citenamefont {Paterova}\ \emph {et~al.}(2018)\citenamefont
  {Paterova}, \citenamefont {Yang}, \citenamefont {An}, \citenamefont
  {Kalashnikov},\ and\ \citenamefont {Krivitsky}}]{Paterova2018NJP}%
  \BibitemOpen
  \bibfield  {author} {\bibinfo {author} {\bibfnamefont {A.}~\bibnamefont
  {Paterova}}, \bibinfo {author} {\bibfnamefont {H.}~\bibnamefont {Yang}},
  \bibinfo {author} {\bibfnamefont {C.}~\bibnamefont {An}}, \bibinfo {author}
  {\bibfnamefont {D.}~\bibnamefont {Kalashnikov}},\ and\ \bibinfo {author}
  {\bibfnamefont {L.}~\bibnamefont {Krivitsky}},\ }\bibfield  {title} {\bibinfo
  {title} {Measurement of infrared optical constants with visible photons},\
  }\href {https://doi.org/10.1088/1367-2630/aab5ce} {\bibfield  {journal}
  {\bibinfo  {journal} {New J. Phys.}\ }\textbf {\bibinfo {volume} {20}},\
  \bibinfo {pages} {043015} (\bibinfo {year} {2018})}\BibitemShut {NoStop}%
\bibitem [{\citenamefont {Grimau~Puigibert}\ \emph {et~al.}(2017)\citenamefont
  {Grimau~Puigibert}, \citenamefont {Aguilar}, \citenamefont {Zhou},
  \citenamefont {Marsili}, \citenamefont {Shaw}, \citenamefont {Verma},
  \citenamefont {Nam}, \citenamefont {Oblak},\ and\ \citenamefont
  {Tittel}}]{Puigibert2017PRL}%
  \BibitemOpen
  \bibfield  {author} {\bibinfo {author} {\bibfnamefont {M.}~\bibnamefont
  {Grimau~Puigibert}}, \bibinfo {author} {\bibfnamefont {G.~H.}\ \bibnamefont
  {Aguilar}}, \bibinfo {author} {\bibfnamefont {Q.}~\bibnamefont {Zhou}},
  \bibinfo {author} {\bibfnamefont {F.}~\bibnamefont {Marsili}}, \bibinfo
  {author} {\bibfnamefont {M.~D.}\ \bibnamefont {Shaw}}, \bibinfo {author}
  {\bibfnamefont {V.~B.}\ \bibnamefont {Verma}}, \bibinfo {author}
  {\bibfnamefont {S.~W.}\ \bibnamefont {Nam}}, \bibinfo {author} {\bibfnamefont
  {D.}~\bibnamefont {Oblak}},\ and\ \bibinfo {author} {\bibfnamefont
  {W.}~\bibnamefont {Tittel}},\ }\bibfield  {title} {\bibinfo {title} {Heralded
  {{Single Photons Based}} on {{Spectral Multiplexing}} and {{Feed-Forward
  Control}}},\ }\href {https://doi.org/10.1103/PhysRevLett.119.083601}
  {\bibfield  {journal} {\bibinfo  {journal} {Phys. Rev. Lett.}\ }\textbf
  {\bibinfo {volume} {119}},\ \bibinfo {pages} {083601} (\bibinfo {year}
  {2017})}\BibitemShut {NoStop}%
\bibitem [{\citenamefont {Joshi}\ \emph {et~al.}(2018)\citenamefont {Joshi},
  \citenamefont {Farsi}, \citenamefont {Clemmen}, \citenamefont {Ramelow},\
  and\ \citenamefont {Gaeta}}]{Joshi2018NC}%
  \BibitemOpen
  \bibfield  {author} {\bibinfo {author} {\bibfnamefont {C.}~\bibnamefont
  {Joshi}}, \bibinfo {author} {\bibfnamefont {A.}~\bibnamefont {Farsi}},
  \bibinfo {author} {\bibfnamefont {S.}~\bibnamefont {Clemmen}}, \bibinfo
  {author} {\bibfnamefont {S.}~\bibnamefont {Ramelow}},\ and\ \bibinfo {author}
  {\bibfnamefont {A.~L.}\ \bibnamefont {Gaeta}},\ }\bibfield  {title} {\bibinfo
  {title} {Frequency multiplexing for quasi-deterministic heralded
  single-photon sources},\ }\href {https://doi.org/10.1038/s41467-018-03254-4}
  {\bibfield  {journal} {\bibinfo  {journal} {Nat Commun}\ }\textbf {\bibinfo
  {volume} {9}},\ \bibinfo {pages} {847} (\bibinfo {year} {2018})}\BibitemShut
  {NoStop}%
\bibitem [{\citenamefont {Olislager}\ \emph {et~al.}(2010)\citenamefont
  {Olislager}, \citenamefont {Cussey}, \citenamefont {Nguyen}, \citenamefont
  {Emplit}, \citenamefont {Massar}, \citenamefont {Merolla},\ and\
  \citenamefont {Huy}}]{Olislager2010PRA}%
  \BibitemOpen
  \bibfield  {author} {\bibinfo {author} {\bibfnamefont {L.}~\bibnamefont
  {Olislager}}, \bibinfo {author} {\bibfnamefont {J.}~\bibnamefont {Cussey}},
  \bibinfo {author} {\bibfnamefont {A.~T.}\ \bibnamefont {Nguyen}}, \bibinfo
  {author} {\bibfnamefont {P.}~\bibnamefont {Emplit}}, \bibinfo {author}
  {\bibfnamefont {S.}~\bibnamefont {Massar}}, \bibinfo {author} {\bibfnamefont
  {J.-M.}\ \bibnamefont {Merolla}},\ and\ \bibinfo {author} {\bibfnamefont
  {K.~P.}\ \bibnamefont {Huy}},\ }\bibfield  {title} {\bibinfo {title}
  {Frequency-bin entangled photons},\ }\href
  {https://doi.org/10.1103/PhysRevA.82.013804} {\bibfield  {journal} {\bibinfo
  {journal} {Phys. Rev. A}\ }\textbf {\bibinfo {volume} {82}},\ \bibinfo
  {pages} {013804} (\bibinfo {year} {2010})}\BibitemShut {NoStop}%
\bibitem [{\citenamefont {Sheridan}\ and\ \citenamefont
  {Scarani}(2010)}]{Sheridan2010PRA}%
  \BibitemOpen
  \bibfield  {author} {\bibinfo {author} {\bibfnamefont {L.}~\bibnamefont
  {Sheridan}}\ and\ \bibinfo {author} {\bibfnamefont {V.}~\bibnamefont
  {Scarani}},\ }\bibfield  {title} {\bibinfo {title} {Security proof for
  quantum key distribution using qudit systems},\ }\href
  {https://doi.org/10.1103/PhysRevA.82.030301} {\bibfield  {journal} {\bibinfo
  {journal} {Phys. Rev. A}\ }\textbf {\bibinfo {volume} {82}},\ \bibinfo
  {pages} {030301} (\bibinfo {year} {2010})}\BibitemShut {NoStop}%
\bibitem [{\citenamefont {Abouraddy}\ \emph {et~al.}(2002)\citenamefont
  {Abouraddy}, \citenamefont {Nasr}, \citenamefont {Saleh}, \citenamefont
  {Sergienko},\ and\ \citenamefont {Teich}}]{Abouraddy2002PRA}%
  \BibitemOpen
  \bibfield  {author} {\bibinfo {author} {\bibfnamefont {A.~F.}\ \bibnamefont
  {Abouraddy}}, \bibinfo {author} {\bibfnamefont {M.~B.}\ \bibnamefont {Nasr}},
  \bibinfo {author} {\bibfnamefont {B.~E.~A.}\ \bibnamefont {Saleh}}, \bibinfo
  {author} {\bibfnamefont {A.~V.}\ \bibnamefont {Sergienko}},\ and\ \bibinfo
  {author} {\bibfnamefont {M.~C.}\ \bibnamefont {Teich}},\ }\bibfield  {title}
  {\bibinfo {title} {Quantum-optical coherence tomography with dispersion
  cancellation},\ }\href {https://doi.org/10.1103/PhysRevA.65.053817}
  {\bibfield  {journal} {\bibinfo  {journal} {Phys. Rev. A}\ }\textbf {\bibinfo
  {volume} {65}},\ \bibinfo {pages} {053817} (\bibinfo {year}
  {2002})}\BibitemShut {NoStop}%
\bibitem [{\citenamefont {Hayama}\ \emph {et~al.}(2022)\citenamefont {Hayama},
  \citenamefont {Cao}, \citenamefont {Okamoto}, \citenamefont {Suezawa},
  \citenamefont {Okano},\ and\ \citenamefont {Takeuchi}}]{Hayama2022OL}%
  \BibitemOpen
  \bibfield  {author} {\bibinfo {author} {\bibfnamefont {K.}~\bibnamefont
  {Hayama}}, \bibinfo {author} {\bibfnamefont {B.}~\bibnamefont {Cao}},
  \bibinfo {author} {\bibfnamefont {R.}~\bibnamefont {Okamoto}}, \bibinfo
  {author} {\bibfnamefont {S.}~\bibnamefont {Suezawa}}, \bibinfo {author}
  {\bibfnamefont {M.}~\bibnamefont {Okano}},\ and\ \bibinfo {author}
  {\bibfnamefont {S.}~\bibnamefont {Takeuchi}},\ }\bibfield  {title} {\bibinfo
  {title} {High-depth-resolution imaging of dispersive samples using quantum
  optical coherence tomography},\ }\href {https://doi.org/10.1364/OL.469874}
  {\bibfield  {journal} {\bibinfo  {journal} {Opt. Lett.}\ }\textbf {\bibinfo
  {volume} {47}},\ \bibinfo {pages} {4949} (\bibinfo {year}
  {2022})}\BibitemShut {NoStop}%
\bibitem [{\citenamefont {Weis}\ and\ \citenamefont
  {Gaylord}(1985)}]{Weis1985APL}%
  \BibitemOpen
  \bibfield  {author} {\bibinfo {author} {\bibfnamefont {R.~S.}\ \bibnamefont
  {Weis}}\ and\ \bibinfo {author} {\bibfnamefont {T.~K.}\ \bibnamefont
  {Gaylord}},\ }\bibfield  {title} {\bibinfo {title} {Lithium niobate:
  {{Summary}} of physical properties and crystal structure},\ }\href
  {https://doi.org/10.1007/BF00614817} {\bibfield  {journal} {\bibinfo
  {journal} {Applied Physics A Solids and Surfaces}\ }\textbf {\bibinfo
  {volume} {37}},\ \bibinfo {pages} {191} (\bibinfo {year} {1985})}\BibitemShut
  {NoStop}%
\bibitem [{\citenamefont {Xue}\ \emph {et~al.}(2021)\citenamefont {Xue},
  \citenamefont {Niu}, \citenamefont {Liu}, \citenamefont {Duan}, \citenamefont
  {Chen}, \citenamefont {Pan}, \citenamefont {Jia}, \citenamefont {Wang},
  \citenamefont {Liu}, \citenamefont {Zhang}, \citenamefont {Xu}, \citenamefont
  {Zhao}, \citenamefont {Cai}, \citenamefont {Gong}, \citenamefont {Hu},
  \citenamefont {Xie},\ and\ \citenamefont {Zhu}}]{Xue2021PRA}%
  \BibitemOpen
  \bibfield  {author} {\bibinfo {author} {\bibfnamefont {G.-T.}\ \bibnamefont
  {Xue}}, \bibinfo {author} {\bibfnamefont {Y.-F.}\ \bibnamefont {Niu}},
  \bibinfo {author} {\bibfnamefont {X.}~\bibnamefont {Liu}}, \bibinfo {author}
  {\bibfnamefont {J.-C.}\ \bibnamefont {Duan}}, \bibinfo {author}
  {\bibfnamefont {W.}~\bibnamefont {Chen}}, \bibinfo {author} {\bibfnamefont
  {Y.}~\bibnamefont {Pan}}, \bibinfo {author} {\bibfnamefont {K.}~\bibnamefont
  {Jia}}, \bibinfo {author} {\bibfnamefont {X.}~\bibnamefont {Wang}}, \bibinfo
  {author} {\bibfnamefont {H.-Y.}\ \bibnamefont {Liu}}, \bibinfo {author}
  {\bibfnamefont {Y.}~\bibnamefont {Zhang}}, \bibinfo {author} {\bibfnamefont
  {P.}~\bibnamefont {Xu}}, \bibinfo {author} {\bibfnamefont {G.}~\bibnamefont
  {Zhao}}, \bibinfo {author} {\bibfnamefont {X.}~\bibnamefont {Cai}}, \bibinfo
  {author} {\bibfnamefont {Y.-X.}\ \bibnamefont {Gong}}, \bibinfo {author}
  {\bibfnamefont {X.}~\bibnamefont {Hu}}, \bibinfo {author} {\bibfnamefont
  {Z.}~\bibnamefont {Xie}},\ and\ \bibinfo {author} {\bibfnamefont
  {S.}~\bibnamefont {Zhu}},\ }\bibfield  {title} {\bibinfo {title} {Ultrabright
  {{Multiplexed Energy-Time-Entangled Photon Generation}} from {{Lithium
  Niobate}} on {{Insulator Chip}}},\ }\href
  {https://doi.org/10.1103/PhysRevApplied.15.064059} {\bibfield  {journal}
  {\bibinfo  {journal} {Phys. Rev. Applied}\ }\textbf {\bibinfo {volume}
  {15}},\ \bibinfo {pages} {064059} (\bibinfo {year} {2021})}\BibitemShut
  {NoStop}%
\bibitem [{\citenamefont {Javid}\ \emph {et~al.}(2021)\citenamefont {Javid},
  \citenamefont {Ling}, \citenamefont {Staffa}, \citenamefont {Li},
  \citenamefont {He},\ and\ \citenamefont {Lin}}]{Javid2021PRL}%
  \BibitemOpen
  \bibfield  {author} {\bibinfo {author} {\bibfnamefont {U.~A.}\ \bibnamefont
  {Javid}}, \bibinfo {author} {\bibfnamefont {J.}~\bibnamefont {Ling}},
  \bibinfo {author} {\bibfnamefont {J.}~\bibnamefont {Staffa}}, \bibinfo
  {author} {\bibfnamefont {M.}~\bibnamefont {Li}}, \bibinfo {author}
  {\bibfnamefont {Y.}~\bibnamefont {He}},\ and\ \bibinfo {author}
  {\bibfnamefont {Q.}~\bibnamefont {Lin}},\ }\bibfield  {title} {\bibinfo
  {title} {Ultrabroadband {{Entangled Photons}} on a {{Nanophotonic Chip}}},\
  }\href {https://doi.org/10.1103/PhysRevLett.127.183601} {\bibfield  {journal}
  {\bibinfo  {journal} {Phys. Rev. Lett.}\ }\textbf {\bibinfo {volume} {127}},\
  \bibinfo {pages} {183601} (\bibinfo {year} {2021})}\BibitemShut {NoStop}%
\bibitem [{\citenamefont {Fang}\ \emph {et~al.}(2024)\citenamefont {Fang},
  \citenamefont {Wang},\ and\ \citenamefont {Lu}}]{Fang2024OE}%
  \BibitemOpen
  \bibfield  {author} {\bibinfo {author} {\bibfnamefont {X.-X.}\ \bibnamefont
  {Fang}}, \bibinfo {author} {\bibfnamefont {L.}~\bibnamefont {Wang}},\ and\
  \bibinfo {author} {\bibfnamefont {H.}~\bibnamefont {Lu}},\ }\bibfield
  {title} {\bibinfo {title} {Efficient generation of broadband photon pairs in
  shallow-etched lithium niobate nanowaveguides},\ }\href
  {https://doi.org/10.1364/OE.519265} {\bibfield  {journal} {\bibinfo
  {journal} {Optics Express}\ }\textbf {\bibinfo {volume} {32}},\ \bibinfo
  {pages} {22945} (\bibinfo {year} {2024})}\BibitemShut {NoStop}%
\bibitem [{\citenamefont {Li}\ \emph {et~al.}(2017)\citenamefont {Li},
  \citenamefont {Chen}, \citenamefont {Jiang},\ and\ \citenamefont
  {Chen}}]{Li2017OL}%
  \BibitemOpen
  \bibfield  {author} {\bibinfo {author} {\bibfnamefont {G.}~\bibnamefont
  {Li}}, \bibinfo {author} {\bibfnamefont {Y.}~\bibnamefont {Chen}}, \bibinfo
  {author} {\bibfnamefont {H.}~\bibnamefont {Jiang}},\ and\ \bibinfo {author}
  {\bibfnamefont {X.}~\bibnamefont {Chen}},\ }\bibfield  {title} {\bibinfo
  {title} {Broadband sum-frequency generation using d\_33 in periodically poled
  {{LiNbO}}\_3 thin film in the telecommunications band},\ }\href
  {https://doi.org/10.1364/OL.42.000939} {\bibfield  {journal} {\bibinfo
  {journal} {Opt. Lett.}\ }\textbf {\bibinfo {volume} {42}},\ \bibinfo {pages}
  {939} (\bibinfo {year} {2017})}\BibitemShut {NoStop}%
\bibitem [{\citenamefont {Jankowski}\ \emph {et~al.}(2020)\citenamefont
  {Jankowski}, \citenamefont {Langrock}, \citenamefont {Desiatov},
  \citenamefont {Marandi}, \citenamefont {Wang}, \citenamefont {Zhang},
  \citenamefont {Phillips}, \citenamefont {Lon{\v c}ar},\ and\ \citenamefont
  {Fejer}}]{Jankowski2020Optica}%
  \BibitemOpen
  \bibfield  {author} {\bibinfo {author} {\bibfnamefont {M.}~\bibnamefont
  {Jankowski}}, \bibinfo {author} {\bibfnamefont {C.}~\bibnamefont {Langrock}},
  \bibinfo {author} {\bibfnamefont {B.}~\bibnamefont {Desiatov}}, \bibinfo
  {author} {\bibfnamefont {A.}~\bibnamefont {Marandi}}, \bibinfo {author}
  {\bibfnamefont {C.}~\bibnamefont {Wang}}, \bibinfo {author} {\bibfnamefont
  {M.}~\bibnamefont {Zhang}}, \bibinfo {author} {\bibfnamefont {C.~R.}\
  \bibnamefont {Phillips}}, \bibinfo {author} {\bibfnamefont {M.}~\bibnamefont
  {Lon{\v c}ar}},\ and\ \bibinfo {author} {\bibfnamefont {M.~M.}\ \bibnamefont
  {Fejer}},\ }\bibfield  {title} {\bibinfo {title} {Ultrabroadband nonlinear
  optics in nanophotonic periodically poled lithium niobate waveguides},\
  }\href {https://doi.org/10.1364/OPTICA.7.000040} {\bibfield  {journal}
  {\bibinfo  {journal} {Optica}\ }\textbf {\bibinfo {volume} {7}},\ \bibinfo
  {pages} {40} (\bibinfo {year} {2020})}\BibitemShut {NoStop}%
\bibitem [{\citenamefont {Yadav}\ \emph {et~al.}(2022)\citenamefont {Yadav},
  \citenamefont {Venkataraman},\ and\ \citenamefont {Ghosh}}]{Yadav2022OL}%
  \BibitemOpen
  \bibfield  {author} {\bibinfo {author} {\bibfnamefont {V.~K.}\ \bibnamefont
  {Yadav}}, \bibinfo {author} {\bibfnamefont {V.}~\bibnamefont
  {Venkataraman}},\ and\ \bibinfo {author} {\bibfnamefont {J.}~\bibnamefont
  {Ghosh}},\ }\bibfield  {title} {\bibinfo {title} {Broadband telecom photon
  pairs from a fiber-integrated ppln ridge waveguide},\ }\href
  {https://doi.org/10.1364/OL.472045} {\bibfield  {journal} {\bibinfo
  {journal} {Opt. Lett.}\ }\textbf {\bibinfo {volume} {47}},\ \bibinfo {pages}
  {5132} (\bibinfo {year} {2022})}\BibitemShut {NoStop}%
\bibitem [{\citenamefont {Roeder}\ \emph {et~al.}(2024)\citenamefont {Roeder},
  \citenamefont {Gnanavel}, \citenamefont {Pollmann}, \citenamefont {Brecht},
  \citenamefont {Stefszky}, \citenamefont {Padberg}, \citenamefont {Eigner},
  \citenamefont {Silberhorn},\ and\ \citenamefont {Brecht}}]{Roeder2024arXiv}%
  \BibitemOpen
  \bibfield  {author} {\bibinfo {author} {\bibfnamefont {F.}~\bibnamefont
  {Roeder}}, \bibinfo {author} {\bibfnamefont {A.}~\bibnamefont {Gnanavel}},
  \bibinfo {author} {\bibfnamefont {R.}~\bibnamefont {Pollmann}}, \bibinfo
  {author} {\bibfnamefont {O.}~\bibnamefont {Brecht}}, \bibinfo {author}
  {\bibfnamefont {M.}~\bibnamefont {Stefszky}}, \bibinfo {author}
  {\bibfnamefont {L.}~\bibnamefont {Padberg}}, \bibinfo {author} {\bibfnamefont
  {C.}~\bibnamefont {Eigner}}, \bibinfo {author} {\bibfnamefont
  {C.}~\bibnamefont {Silberhorn}},\ and\ \bibinfo {author} {\bibfnamefont
  {B.}~\bibnamefont {Brecht}},\ }\href {https://arxiv.org/abs/2408.12203}
  {\bibinfo {title} {Ultra-broadband non-degenerate guided-wave bi-photon
  source in the near and mid-infrared}} (\bibinfo {year} {2024}),\ \Eprint
  {https://arxiv.org/abs/2408.12203} {arXiv:2408.12203 [quant-ph]} \BibitemShut
  {NoStop}%
\bibitem [{\citenamefont {Shaked}\ \emph {et~al.}(2014)\citenamefont {Shaked},
  \citenamefont {Pomerantz}, \citenamefont {Vered},\ and\ \citenamefont
  {Pe\'er}}]{Shaked2014NJP}%
  \BibitemOpen
  \bibfield  {author} {\bibinfo {author} {\bibfnamefont {Y.}~\bibnamefont
  {Shaked}}, \bibinfo {author} {\bibfnamefont {R.}~\bibnamefont {Pomerantz}},
  \bibinfo {author} {\bibfnamefont {R.~Z.}\ \bibnamefont {Vered}},\ and\
  \bibinfo {author} {\bibfnamefont {A.}~\bibnamefont {Pe\'er}},\ }\bibfield
  {title} {\bibinfo {title} {Observing the nonclassical nature of
  ultra-broadband bi-photons at ultrafast speed},\ }\href
  {https://doi.org/10.1088/1367-2630/16/5/053012} {\bibfield  {journal}
  {\bibinfo  {journal} {New Journal of Physics}\ }\textbf {\bibinfo {volume}
  {16}},\ \bibinfo {pages} {053012} (\bibinfo {year} {2014})}\BibitemShut
  {NoStop}%
\bibitem [{\citenamefont {Nasr}\ \emph {et~al.}(2008)\citenamefont {Nasr},
  \citenamefont {Carrasco}, \citenamefont {Saleh}, \citenamefont {Sergienko},
  \citenamefont {Teich}, \citenamefont {Torres}, \citenamefont {Torner},
  \citenamefont {Hum},\ and\ \citenamefont {Fejer}}]{Nasr2008PRL}%
  \BibitemOpen
  \bibfield  {author} {\bibinfo {author} {\bibfnamefont {M.~B.}\ \bibnamefont
  {Nasr}}, \bibinfo {author} {\bibfnamefont {S.}~\bibnamefont {Carrasco}},
  \bibinfo {author} {\bibfnamefont {B.~E.~A.}\ \bibnamefont {Saleh}}, \bibinfo
  {author} {\bibfnamefont {A.~V.}\ \bibnamefont {Sergienko}}, \bibinfo {author}
  {\bibfnamefont {M.~C.}\ \bibnamefont {Teich}}, \bibinfo {author}
  {\bibfnamefont {J.~P.}\ \bibnamefont {Torres}}, \bibinfo {author}
  {\bibfnamefont {L.}~\bibnamefont {Torner}}, \bibinfo {author} {\bibfnamefont
  {D.~S.}\ \bibnamefont {Hum}},\ and\ \bibinfo {author} {\bibfnamefont {M.~M.}\
  \bibnamefont {Fejer}},\ }\bibfield  {title} {\bibinfo {title} {Ultrabroadband
  biphotons generated via chirped quasi-phase-matched optical parametric
  down-conversion},\ }\href {https://doi.org/10.1103/PhysRevLett.100.183601}
  {\bibfield  {journal} {\bibinfo  {journal} {Phys. Rev. Lett.}\ }\textbf
  {\bibinfo {volume} {100}},\ \bibinfo {pages} {183601} (\bibinfo {year}
  {2008})}\BibitemShut {NoStop}%
\bibitem [{\citenamefont {Tanaka}\ \emph {et~al.}(2012)\citenamefont {Tanaka},
  \citenamefont {Okamoto}, \citenamefont {Lim}, \citenamefont {Subashchandran},
  \citenamefont {Okano}, \citenamefont {Zhang}, \citenamefont {Kang},
  \citenamefont {Chen}, \citenamefont {Wu}, \citenamefont {Hirohata},
  \citenamefont {Kurimura},\ and\ \citenamefont {Takeuchi}}]{Tanaka2012OE}%
  \BibitemOpen
  \bibfield  {author} {\bibinfo {author} {\bibfnamefont {A.}~\bibnamefont
  {Tanaka}}, \bibinfo {author} {\bibfnamefont {R.}~\bibnamefont {Okamoto}},
  \bibinfo {author} {\bibfnamefont {H.~H.}\ \bibnamefont {Lim}}, \bibinfo
  {author} {\bibfnamefont {S.}~\bibnamefont {Subashchandran}}, \bibinfo
  {author} {\bibfnamefont {M.}~\bibnamefont {Okano}}, \bibinfo {author}
  {\bibfnamefont {L.}~\bibnamefont {Zhang}}, \bibinfo {author} {\bibfnamefont
  {L.}~\bibnamefont {Kang}}, \bibinfo {author} {\bibfnamefont {J.}~\bibnamefont
  {Chen}}, \bibinfo {author} {\bibfnamefont {P.}~\bibnamefont {Wu}}, \bibinfo
  {author} {\bibfnamefont {T.}~\bibnamefont {Hirohata}}, \bibinfo {author}
  {\bibfnamefont {S.}~\bibnamefont {Kurimura}},\ and\ \bibinfo {author}
  {\bibfnamefont {S.}~\bibnamefont {Takeuchi}},\ }\bibfield  {title} {\bibinfo
  {title} {Noncollinear parametric fluorescence by chirped quasi-phase matching
  for monocycle temporal entanglement},\ }\href
  {https://doi.org/10.1364/OE.20.025228} {\bibfield  {journal} {\bibinfo
  {journal} {Opt. Express}\ }\textbf {\bibinfo {volume} {20}},\ \bibinfo
  {pages} {25228} (\bibinfo {year} {2012})}\BibitemShut {NoStop}%
\bibitem [{\citenamefont {Cao}\ \emph {et~al.}(2021)\citenamefont {Cao},
  \citenamefont {Hisamitsu}, \citenamefont {Tokuda}, \citenamefont {Kurimura},
  \citenamefont {Okamoto},\ and\ \citenamefont {Takeuchi}}]{Cao2021OE}%
  \BibitemOpen
  \bibfield  {author} {\bibinfo {author} {\bibfnamefont {B.}~\bibnamefont
  {Cao}}, \bibinfo {author} {\bibfnamefont {M.}~\bibnamefont {Hisamitsu}},
  \bibinfo {author} {\bibfnamefont {K.}~\bibnamefont {Tokuda}}, \bibinfo
  {author} {\bibfnamefont {S.}~\bibnamefont {Kurimura}}, \bibinfo {author}
  {\bibfnamefont {R.}~\bibnamefont {Okamoto}},\ and\ \bibinfo {author}
  {\bibfnamefont {S.}~\bibnamefont {Takeuchi}},\ }\bibfield  {title} {\bibinfo
  {title} {Efficient generation of ultra-broadband parametric fluorescence
  using chirped quasi-phase-matched waveguide devices},\ }\href
  {https://doi.org/10.1364/OE.426575} {\bibfield  {journal} {\bibinfo
  {journal} {Opt. Express}\ }\textbf {\bibinfo {volume} {29}},\ \bibinfo
  {pages} {21615} (\bibinfo {year} {2021})}\BibitemShut {NoStop}%
\bibitem [{\citenamefont {Hojo}\ and\ \citenamefont
  {Tanaka}(2021)}]{Hojo2021SR}%
  \BibitemOpen
  \bibfield  {author} {\bibinfo {author} {\bibfnamefont {M.}~\bibnamefont
  {Hojo}}\ and\ \bibinfo {author} {\bibfnamefont {K.}~\bibnamefont {Tanaka}},\
  }\bibfield  {title} {\bibinfo {title} {Broadband infrared light source by
  simultaneous parametric down-conversion},\ }\href
  {https://doi.org/10.1038/s41598-021-97531-w} {\bibfield  {journal} {\bibinfo
  {journal} {Scientific Reports}\ }\textbf {\bibinfo {volume} {11}},\ \bibinfo
  {pages} {17986} (\bibinfo {year} {2021})}\BibitemShut {NoStop}%
\bibitem [{\citenamefont {Cao}\ \emph {et~al.}(2023)\citenamefont {Cao},
  \citenamefont {Hayama}, \citenamefont {Suezawa}, \citenamefont {Hisamitsu},
  \citenamefont {Tokuda}, \citenamefont {Kurimura}, \citenamefont {Okamoto},\
  and\ \citenamefont {Takeuchi}}]{Cao2023OE}%
  \BibitemOpen
  \bibfield  {author} {\bibinfo {author} {\bibfnamefont {B.}~\bibnamefont
  {Cao}}, \bibinfo {author} {\bibfnamefont {K.}~\bibnamefont {Hayama}},
  \bibinfo {author} {\bibfnamefont {S.}~\bibnamefont {Suezawa}}, \bibinfo
  {author} {\bibfnamefont {M.}~\bibnamefont {Hisamitsu}}, \bibinfo {author}
  {\bibfnamefont {K.}~\bibnamefont {Tokuda}}, \bibinfo {author} {\bibfnamefont
  {S.}~\bibnamefont {Kurimura}}, \bibinfo {author} {\bibfnamefont
  {R.}~\bibnamefont {Okamoto}},\ and\ \bibinfo {author} {\bibfnamefont
  {S.}~\bibnamefont {Takeuchi}},\ }\bibfield  {title} {\bibinfo {title}
  {Non-collinear generation of ultra-broadband parametric fluorescence photon
  pairs using chirped quasi-phase matching slab waveguides},\ }\href
  {https://doi.org/10.1364/OE.488978} {\bibfield  {journal} {\bibinfo
  {journal} {Optics Express}\ }\textbf {\bibinfo {volume} {31}},\ \bibinfo
  {pages} {23551} (\bibinfo {year} {2023})}\BibitemShut {NoStop}%
\bibitem [{\citenamefont {Tashima}\ \emph {et~al.}(2024)\citenamefont
  {Tashima}, \citenamefont {Mukai}, \citenamefont {Arahata}, \citenamefont
  {Oda}, \citenamefont {Hisamitsu}, \citenamefont {Tokuda}, \citenamefont
  {Okamoto},\ and\ \citenamefont {Takeuchi}}]{Tashima2024Optica}%
  \BibitemOpen
  \bibfield  {author} {\bibinfo {author} {\bibfnamefont {T.}~\bibnamefont
  {Tashima}}, \bibinfo {author} {\bibfnamefont {Y.}~\bibnamefont {Mukai}},
  \bibinfo {author} {\bibfnamefont {M.}~\bibnamefont {Arahata}}, \bibinfo
  {author} {\bibfnamefont {N.}~\bibnamefont {Oda}}, \bibinfo {author}
  {\bibfnamefont {M.}~\bibnamefont {Hisamitsu}}, \bibinfo {author}
  {\bibfnamefont {K.}~\bibnamefont {Tokuda}}, \bibinfo {author} {\bibfnamefont
  {R.}~\bibnamefont {Okamoto}},\ and\ \bibinfo {author} {\bibfnamefont
  {S.}~\bibnamefont {Takeuchi}},\ }\bibfield  {title} {\bibinfo {title}
  {Ultra-broadband quantum infrared spectroscopy},\ }\href
  {https://doi.org/10.1364/OPTICA.504450} {\bibfield  {journal} {\bibinfo
  {journal} {Optica}\ }\textbf {\bibinfo {volume} {11}},\ \bibinfo {pages} {81}
  (\bibinfo {year} {2024})}\BibitemShut {NoStop}%
\bibitem [{\citenamefont {Harris}(2007)}]{Harris2007PRL}%
  \BibitemOpen
  \bibfield  {author} {\bibinfo {author} {\bibfnamefont {S.~E.}\ \bibnamefont
  {Harris}},\ }\bibfield  {title} {\bibinfo {title} {Chirp and compress: Toward
  single-cycle biphotons},\ }\href
  {https://doi.org/10.1103/PhysRevLett.98.063602} {\bibfield  {journal}
  {\bibinfo  {journal} {Phys. Rev. Lett.}\ }\textbf {\bibinfo {volume} {98}},\
  \bibinfo {pages} {063602} (\bibinfo {year} {2007})}\BibitemShut {NoStop}%
\bibitem [{\citenamefont {Zhu}\ and\ \citenamefont {Jin}(2025)}]{Zhu2025APL}%
  \BibitemOpen
  \bibfield  {author} {\bibinfo {author} {\bibfnamefont {W.-X.}\ \bibnamefont
  {Zhu}}\ and\ \bibinfo {author} {\bibfnamefont {R.-B.}\ \bibnamefont {Jin}},\
  }\bibfield  {title} {\bibinfo {title} {4780 nm ultra-broadband entangled
  biphotons from a chirped {{PPLN}}},\ }\href
  {https://doi.org/10.1063/5.0237968} {\bibfield  {journal} {\bibinfo
  {journal} {Applied Physics Letters}\ }\textbf {\bibinfo {volume} {126}},\
  \bibinfo {pages} {014001} (\bibinfo {year} {2025})}\BibitemShut {NoStop}%
\bibitem [{\citenamefont {Chen}\ \emph {et~al.}(2014)\citenamefont {Chen},
  \citenamefont {Ren}, \citenamefont {Liu}, \citenamefont {Zhang},
  \citenamefont {Sheng}, \citenamefont {Ma},\ and\ \citenamefont
  {Li}}]{Chen2014Light}%
  \BibitemOpen
  \bibfield  {author} {\bibinfo {author} {\bibfnamefont {B.-Q.}\ \bibnamefont
  {Chen}}, \bibinfo {author} {\bibfnamefont {M.-L.}\ \bibnamefont {Ren}},
  \bibinfo {author} {\bibfnamefont {R.-J.}\ \bibnamefont {Liu}}, \bibinfo
  {author} {\bibfnamefont {C.}~\bibnamefont {Zhang}}, \bibinfo {author}
  {\bibfnamefont {Y.}~\bibnamefont {Sheng}}, \bibinfo {author} {\bibfnamefont
  {B.-Q.}\ \bibnamefont {Ma}},\ and\ \bibinfo {author} {\bibfnamefont {Z.-Y.}\
  \bibnamefont {Li}},\ }\bibfield  {title} {\bibinfo {title} {Simultaneous
  broadband generation of second and third harmonics from chirped nonlinear
  photonic crystals},\ }\href {https://doi.org/10.1038/lsa.2014.70} {\bibfield
  {journal} {\bibinfo  {journal} {Light: Science \& Applications}\ }\textbf
  {\bibinfo {volume} {3}},\ \bibinfo {pages} {e189} (\bibinfo {year}
  {2014})}\BibitemShut {NoStop}%
\bibitem [{\citenamefont {Chen}\ \emph {et~al.}(2021)\citenamefont {Chen},
  \citenamefont {Hong}, \citenamefont {Hu},\ and\ \citenamefont
  {Li}}]{Chen2021Research}%
  \BibitemOpen
  \bibfield  {author} {\bibinfo {author} {\bibfnamefont {B.}~\bibnamefont
  {Chen}}, \bibinfo {author} {\bibfnamefont {L.}~\bibnamefont {Hong}}, \bibinfo
  {author} {\bibfnamefont {C.}~\bibnamefont {Hu}},\ and\ \bibinfo {author}
  {\bibfnamefont {Z.}~\bibnamefont {Li}},\ }\bibfield  {title} {\bibinfo
  {title} {White laser realized via synergic second- and third-order
  nonlinearities},\ }\bibfield  {journal} {\bibinfo  {journal} {Research}\
  }\textbf {\bibinfo {volume} {2021}},\ \href
  {https://doi.org/10.34133/2021/1539730} {10.34133/2021/1539730} (\bibinfo
  {year} {2021}),\ \Eprint
  {https://arxiv.org/abs/https://spj.science.org/doi/pdf/10.34133/2021/1539730}
  {https://spj.science.org/doi/pdf/10.34133/2021/1539730} \BibitemShut
  {NoStop}%
\bibitem [{\citenamefont {Wu}\ \emph {et~al.}(2022)\citenamefont {Wu},
  \citenamefont {Zhang}, \citenamefont {Hao}, \citenamefont {Zhang},
  \citenamefont {Ma}, \citenamefont {Bo}, \citenamefont {Zhang},\ and\
  \citenamefont {Xu}}]{Wu2022OL}%
  \BibitemOpen
  \bibfield  {author} {\bibinfo {author} {\bibfnamefont {X.}~\bibnamefont
  {Wu}}, \bibinfo {author} {\bibfnamefont {L.}~\bibnamefont {Zhang}}, \bibinfo
  {author} {\bibfnamefont {Z.}~\bibnamefont {Hao}}, \bibinfo {author}
  {\bibfnamefont {R.}~\bibnamefont {Zhang}}, \bibinfo {author} {\bibfnamefont
  {R.}~\bibnamefont {Ma}}, \bibinfo {author} {\bibfnamefont {F.}~\bibnamefont
  {Bo}}, \bibinfo {author} {\bibfnamefont {G.}~\bibnamefont {Zhang}},\ and\
  \bibinfo {author} {\bibfnamefont {J.}~\bibnamefont {Xu}},\ }\bibfield
  {title} {\bibinfo {title} {Broadband second-harmonic generation in
  step-chirped periodically poled lithium niobate waveguides},\ }\href
  {https://doi.org/10.1364/OL.450547} {\bibfield  {journal} {\bibinfo
  {journal} {Opt. Lett., OL}\ }\textbf {\bibinfo {volume} {47}},\ \bibinfo
  {pages} {1574} (\bibinfo {year} {2022})}\BibitemShut {NoStop}%
\bibitem [{\citenamefont {Zhang}\ \emph {et~al.}(2022)\citenamefont {Zhang},
  \citenamefont {Li}, \citenamefont {Zhu}, \citenamefont {Cai},\ and\
  \citenamefont {Hu}}]{Zhang2022OME}%
  \BibitemOpen
  \bibfield  {author} {\bibinfo {author} {\bibfnamefont {H.}~\bibnamefont
  {Zhang}}, \bibinfo {author} {\bibfnamefont {Q.}~\bibnamefont {Li}}, \bibinfo
  {author} {\bibfnamefont {H.}~\bibnamefont {Zhu}}, \bibinfo {author}
  {\bibfnamefont {L.}~\bibnamefont {Cai}},\ and\ \bibinfo {author}
  {\bibfnamefont {H.}~\bibnamefont {Hu}},\ }\bibfield  {title} {\bibinfo
  {title} {Second harmonic generation by quasi-phase matching in a lithium
  niobate thin film},\ }\href {https://doi.org/10.1364/OME.452483} {\bibfield
  {journal} {\bibinfo  {journal} {Opt. Mater. Express}\ }\textbf {\bibinfo
  {volume} {12}},\ \bibinfo {pages} {2252} (\bibinfo {year}
  {2022})}\BibitemShut {NoStop}%
\bibitem [{\citenamefont {Hong}\ \emph {et~al.}(1987)\citenamefont {Hong},
  \citenamefont {Ou},\ and\ \citenamefont {Mandel}}]{Hong1987PRL}%
  \BibitemOpen
  \bibfield  {author} {\bibinfo {author} {\bibfnamefont {C.~K.}\ \bibnamefont
  {Hong}}, \bibinfo {author} {\bibfnamefont {Z.~Y.}\ \bibnamefont {Ou}},\ and\
  \bibinfo {author} {\bibfnamefont {L.}~\bibnamefont {Mandel}},\ }\bibfield
  {title} {\bibinfo {title} {Measurement of subpicosecond time intervals
  between two photons by interference},\ }\href
  {https://doi.org/10.1103/PhysRevLett.59.2044} {\bibfield  {journal} {\bibinfo
   {journal} {Phys. Rev. Lett.}\ }\textbf {\bibinfo {volume} {59}},\ \bibinfo
  {pages} {2044} (\bibinfo {year} {1987})}\BibitemShut {NoStop}%
\bibitem [{\citenamefont {Lyons}\ \emph {et~al.}(2018)\citenamefont {Lyons},
  \citenamefont {Knee}, \citenamefont {Bolduc}, \citenamefont {Roger},
  \citenamefont {Leach}, \citenamefont {Gauger},\ and\ \citenamefont
  {Faccio}}]{Lyons2018SA}%
  \BibitemOpen
  \bibfield  {author} {\bibinfo {author} {\bibfnamefont {A.}~\bibnamefont
  {Lyons}}, \bibinfo {author} {\bibfnamefont {G.~C.}\ \bibnamefont {Knee}},
  \bibinfo {author} {\bibfnamefont {E.}~\bibnamefont {Bolduc}}, \bibinfo
  {author} {\bibfnamefont {T.}~\bibnamefont {Roger}}, \bibinfo {author}
  {\bibfnamefont {J.}~\bibnamefont {Leach}}, \bibinfo {author} {\bibfnamefont
  {E.~M.}\ \bibnamefont {Gauger}},\ and\ \bibinfo {author} {\bibfnamefont
  {D.}~\bibnamefont {Faccio}},\ }\bibfield  {title} {\bibinfo {title}
  {Attosecond-resolution hong-ou-mandel interferometry},\ }\href
  {https://doi.org/10.1126/sciadv.aap9416} {\bibfield  {journal} {\bibinfo
  {journal} {Science Advances}\ }\textbf {\bibinfo {volume} {4}},\ \bibinfo
  {pages} {eaap9416} (\bibinfo {year} {2018})},\ \Eprint
  {https://arxiv.org/abs/https://www.science.org/doi/pdf/10.1126/sciadv.aap9416}
  {https://www.science.org/doi/pdf/10.1126/sciadv.aap9416} \BibitemShut
  {NoStop}%
\bibitem [{\citenamefont {Guo}\ \emph {et~al.}(2023)\citenamefont {Guo},
  \citenamefont {Yang}, \citenamefont {Zeng}, \citenamefont {Ding},
  \citenamefont {Shimizu},\ and\ \citenamefont {Jin}}]{Guo2023OE}%
  \BibitemOpen
  \bibfield  {author} {\bibinfo {author} {\bibfnamefont {Y.}~\bibnamefont
  {Guo}}, \bibinfo {author} {\bibfnamefont {Z.-X.}\ \bibnamefont {Yang}},
  \bibinfo {author} {\bibfnamefont {Z.-Q.}\ \bibnamefont {Zeng}}, \bibinfo
  {author} {\bibfnamefont {C.}~\bibnamefont {Ding}}, \bibinfo {author}
  {\bibfnamefont {R.}~\bibnamefont {Shimizu}},\ and\ \bibinfo {author}
  {\bibfnamefont {R.-B.}\ \bibnamefont {Jin}},\ }\bibfield  {title} {\bibinfo
  {title} {Comparison of multi-mode hong-ou-mandel interference and multi-slit
  interference},\ }\href {https://doi.org/10.1364/OE.501645} {\bibfield
  {journal} {\bibinfo  {journal} {Opt. Express}\ }\textbf {\bibinfo {volume}
  {31}},\ \bibinfo {pages} {32849} (\bibinfo {year} {2023})}\BibitemShut
  {NoStop}%
\bibitem [{\citenamefont {Jin}\ \emph {et~al.}(2024)\citenamefont {Jin},
  \citenamefont {Zeng}, \citenamefont {You},\ and\ \citenamefont
  {Yuan}}]{Jin2024PQE}%
  \BibitemOpen
  \bibfield  {author} {\bibinfo {author} {\bibfnamefont {R.-B.}\ \bibnamefont
  {Jin}}, \bibinfo {author} {\bibfnamefont {Z.-Q.}\ \bibnamefont {Zeng}},
  \bibinfo {author} {\bibfnamefont {C.}~\bibnamefont {You}},\ and\ \bibinfo
  {author} {\bibfnamefont {C.}~\bibnamefont {Yuan}},\ }\bibfield  {title}
  {\bibinfo {title} {Quantum interferometers: Principles and applications},\
  }\href {https://doi.org/https://doi.org/10.1016/j.pquantelec.2024.100519}
  {\bibfield  {journal} {\bibinfo  {journal} {Progress in Quantum Electronics}\
  }\textbf {\bibinfo {volume} {96}},\ \bibinfo {pages} {100519} (\bibinfo
  {year} {2024})}\BibitemShut {NoStop}%
\end{thebibliography}%
\end{document}